\newcounter{myctr}
\def\myitem{\refstepcounter{myctr}\bibfont\noindent\ifnum\themyctr>9\else\phantom{0}\fi\hangindent17pt\themyctr.\enskip}

\documentclass{ws-ijqi}

\usepackage{graphicx}
\usepackage{hyperref}
\usepackage{amsmath}
\usepackage{mathrsfs}
\usepackage{amsfonts}
\usepackage{bm} 
\usepackage{subfigure}
\usepackage{epsfig} 
\usepackage{epstopdf}
\usepackage{float} 
\usepackage{ bbold }
\usepackage{tikz}
\usetikzlibrary{shapes.arrows}
\usepackage[percent]{overpic}
\usepackage{comment}
\usepackage{pstricks}
\usepackage{filecontents}
\usepackage{enumitem}  
\usepackage{zref-xr}
\usepackage[utf8]{inputenc}

\hypersetup{
  colorlinks,
  citecolor=blue	,
  linkcolor=blue	,
  urlcolor=blue	}

\begin{document}

\title{ QUANTUM KEY DISTRIBUTION BASED ON THE QUANTUM ERASER}

\author{TAREK A. ELSAYED}

\address{Department of Physics, School of Science and Engineering, The American University in Cairo, AUC Avenue, P.O. Box 74, New Cairo, 11835, Egypt}
\address{Zewail City of Science and Technology, 6th of October City, Giza 12578, Egypt}
\address{Department of Physics and Astronomy, Hunter College of the City University of New York,New York 10065, USA\\
tarek.ahmed.elsayed@gmail.com
}

\maketitle

\begin{history}
\end{history}



\begin{abstract}
Quantum information and quantum foundations are becoming popular topics for advanced undergraduate courses. Many of the fundamental concepts and applications in these two fields, such as delayed choice experiments and quantum encryption, are comprehensible to undergraduates with basic knowledge of quantum mechanics. In this paper, we show that the quantum eraser, usually used to study the duality between wave and particle properties, can also serve as a generic platform for quantum key distribution. We present a pedagogical example of an algorithm to securely share random keys using the quantum eraser platform and propose its implementation with quantum circuits.  
 
\end{abstract}

\section{Introduction}
The fields of quantum foundations and quantum information are very good examples of themes that are being actively investigated at the research level and yet are accessible to  undergraduate students who completed a first course on quantum mechanics\cite{strauch2016,ashby2016,ferrari2010,courtney2020}. Indeed, many undergraduate  curricula include courses and seminars on topics such as quantum communication, quantum computing and algorithms, entanglement, quantum control, ...etc. The two fields are strongly interrelated and the advances in either one leads to breaking new grounds in the other one\cite{barnum2018}.  The purpose of this paper is to provide a pedagogical example of a quantum communication application motivated by foundational aspects of quantum mechanics that illustrates the connections between the two subjects.

The principle of complementarity in quantum systems has been one of the cornerstone ideas in quantum mechanics since its inception in the 1920s. In short, this principle states that quantum systems cannot exhibit full wave and particle properties at the same time. For example, in the double slit experiment, one cannot pinpoint the slit through which the quantum particle  has passed  (the path) and at the same time maintain the interference fringes on the screen. Very important variations of this experiment such as delaying the choice of whether or not to measure the path until after the particle has passed through the double slits (delayed choice experiments\cite{wheeler1978})  or reversing the choice, i.e., erasing the which-way information (WWI)\cite{scully1982}, have been implemented in various experiments over the last two decades\cite{kim2000,walborn2002,jacques2007,manning2015,ma2016}.  

On the other hand, quantum key distribution (QKD) has been one of the earliest applications of quantum information\cite{gisin2002,scarani2009}. A QKD algorithm aims at securely sharing a random set of bits (the key) between two distant parties. The most famous QKD algorithms are the Bennett-Brassard 1984 (BB84) protocol which uses single photons\cite{bb84} and the Ekert 1991 (Ekert91) protocol\cite{ekert91} which uses pairs of entangled photons.  
The two fields of quantum information and quantum foundations are strongly related and are often pursued in parallel by many groups. Recently, new protocols for quantum communications and quantum computing based on notions in the foundations of quantum mechanics such as counterfactuality and interaction-free measurement  have been proposed\cite{mitchison2001,salih2013,cao2017,liu2017}. A core ingredient in QKD algorithms is the existence of two non-orthogonal encoding schemes for representing the logical bits.  In this paper, we show that the quantum eraser platform can provide such encoding schemes. Specifically, we propose new encryption schemes qualitatively similar to the BB84 and Ekert91 protocols and based on the single photon and two-photon quantum eraser platforms respectively. We also illustrate how to implement the proposed schemes using the available quantum circuit components.


 \begin{figure*}[!t] \setlength{\unitlength}{0.1cm}

\begin{picture}(110 , 45 ) 
{
\put(0, 0)  {\includegraphics[ width=4.3cm,height=4.cm]{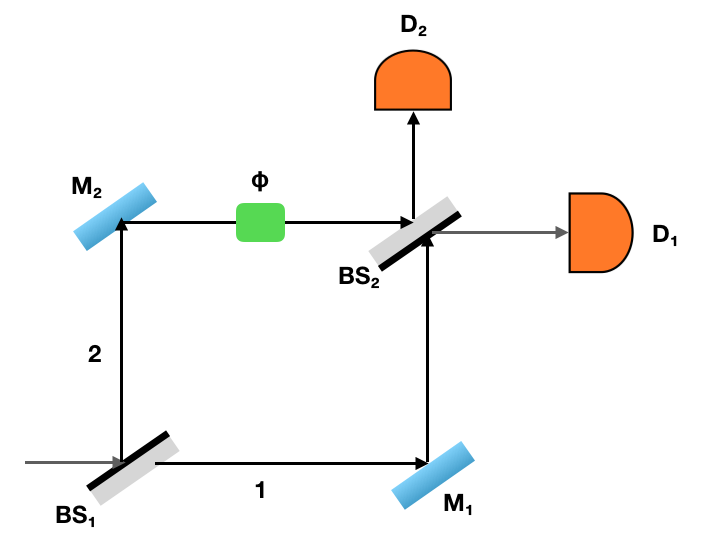}}
\put(43, 0)  {\includegraphics[ width=4.3cm,height=4.cm]{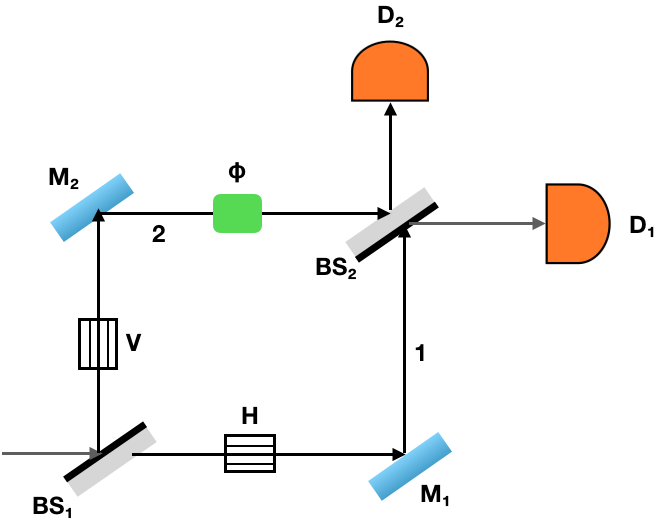}}
\put(86, 0)  {\includegraphics[ width=4.3cm,height=4.cm]{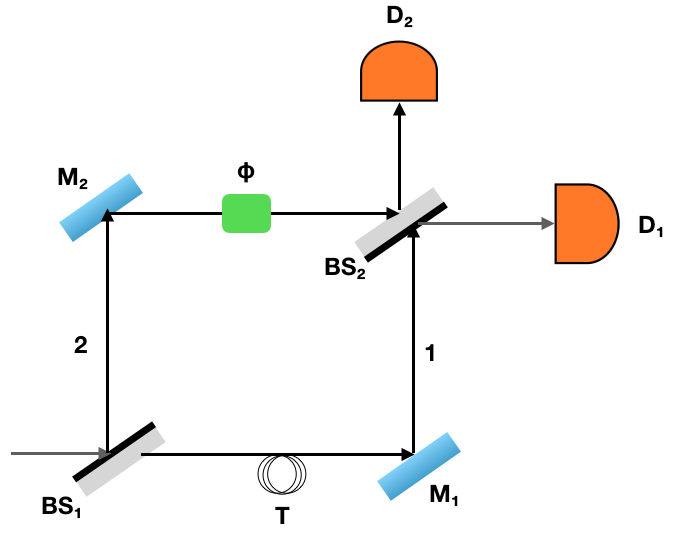}}

\put(20,43) {\small \bf (a) }
\put(55,43) {\small \bf (b) }
\put(100,43) {\small \bf (c) }

\put(73,21) {\tiny \bf A }
\put(69,28) {\tiny \bf B }

\put(30,21) {\tiny \bf A }
\put(26,28) {\tiny \bf B }

\put(116,21) {\tiny \bf A }
\put(112,28) {\tiny \bf B }

}
\end{picture} 

\caption{ \label{mzi} (a) Mach-Zender interferometer with phase difference $\phi$ between the two paths. The detection probability at detectors $\text{D}_1$ and $\text{D}_2$ exhibit sinusoidal dependence on $\phi$ due to the interference between the two paths.  (b,c) Which-way information is added to the setup. In (b),  a source of unpolarized photons is used and vertical and horizontal polarizers are added to the two paths, while in (c), a time delay is added to one path.}
\end{figure*}
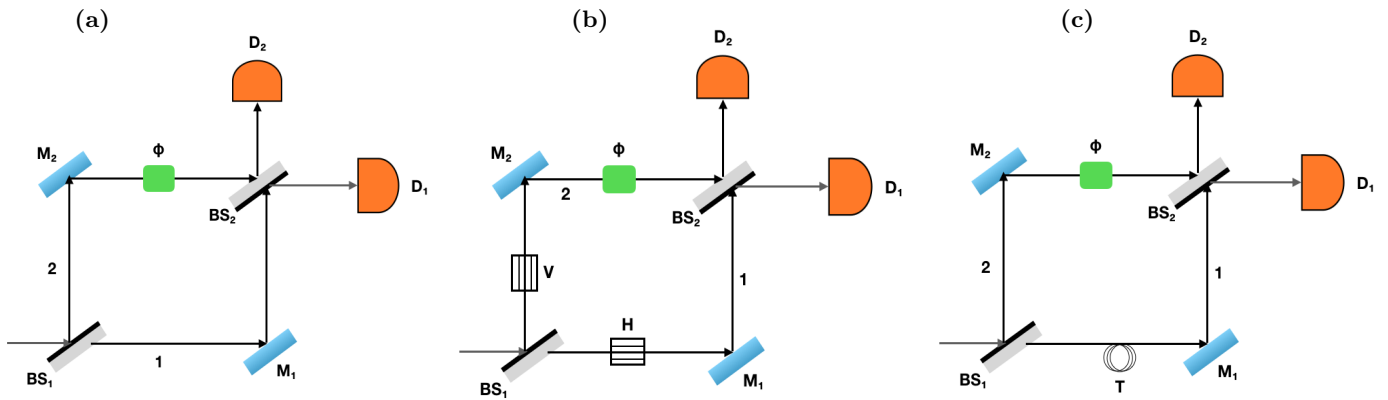 

 \normalsize

\section{The different flavors of the quantum eraser}
First, we give an overview of the different flavors of the single photon and the two-photon quantum erasers. In both schemes, the Mach-Zender interferometer (MZI) will be a fundamental building block. In MZI, a photon is sent through a beam splitter (BS), that allows its wavefunction to become in a superposition of the two paths 1 and 2 that recombine at another beam splitter before the photon is detected by either detector $\text{D}_1$ or $\text{D}_2$ (see Fig. 1-a).

Inside the interferometer, the state of the photon is expressed as $\psi=\frac{1}{\sqrt{2}} ( |1\rangle+ |2\rangle)$, where $|1\rangle$ and $|2\rangle$ represent the states of the photon in the two paths. If we impose a phase difference $\phi$ between the two paths and let the paths emerging towards the two detectors be labeled A and B, the wavefunction after the second beam splitter is $ \frac{1}{2} ( (1+e^{i\phi}) |A\rangle- (1-e^{i\phi})|B\rangle)$. Consequently, the probability of detecting the photon in $\text{D}_1$ is $P_1(\phi)=\cos^2(\phi/2)$ while the probability of detecting the photon in $\text{D}_2$ is $P_2(\phi)=\sin^2(\phi/2)$. This sinusoidal dependence on $\phi$ is the hallmark of the interference pattern of the Mach-Zender interferometer.

In order to know through which path the photon has traveled, the two paths should be given two different labels through a new degree of freedom of the same photon or another photon. Consider the case when the new degree of freedom pertains to the same photon and let us call the states corresponding to these new labels $|H\rangle$ and $|V\rangle$. The state of the photon after emerging from the second beam splitter is thus 
$\frac{1}{2} |A \rangle \otimes \left \{ |H\rangle+e^{i \phi}|V\rangle  \right \}-\frac{1}{2} |B \rangle \otimes \left \{ |H\rangle-e^{i \phi}|V\rangle  \right \}$.  It it clear that when $|H\rangle$ and $|V\rangle$ are orthogonal, i.e., the two paths are fully distinguishable, the probability of detecting the photon at any of the two detectors will be $0.5$, i.e., the interference pattern vanishes. This can be achieved, for example, using the polarization of the photon (i.e., vertical or horizontal polarization; see Fig. \ref{mzi}-b), or the time of arrival of the photon (i.e., by introducing a sufficiently large time delay T in one path; see Fig. \ref{mzi}-c) as in time-bin qubits\cite{marcikic2004}. The former case can be implemented using a source of unpolarized photons, and absorbing polarizers in each path, or a source of linearly polarized photons with a 45° half-wave plate (HWP) in one path only that rotates the plane of polarization by 90 degrees. In the later case (Fig. \ref{mzi}-c), the time delay should be adjusted in a way that does not alter the phase difference between the two paths. 

The quantum eraser setup aims at `erasing' the which-way information after the photon has passed through $\text{BS}_2$, thus restoring the sinusoidal interference pattern at $\text{D}_1$ and $\text{D}_2$. This can be done for example by exploiting the correlation between the path information and the other degree of freedom and sorting the measurement outcomes of $\text{D}_1$ and $\text{D}_2$ into sub-ensembles that exhibit complementary interference patterns (fringes and anti-fringes)\cite{englert2000,kwiat2004}. In the two-photon quantum eraser experiments, the WWI is correlated with a quantum degree of freedom carried by another particle, called the idler photon, where sorting or coincidence counting is performed according to the measurement outcomes of that photon. The ability to erase WWI and restore the interference pattern after the photon has left the interferometer illustrates the  complementary nature of the wave and particle properties of the photons and has invoked an interesting discussion in the foundations of quantum mechanics that still stimulates new models and experiments\cite{chaves2018,qin2018}.



 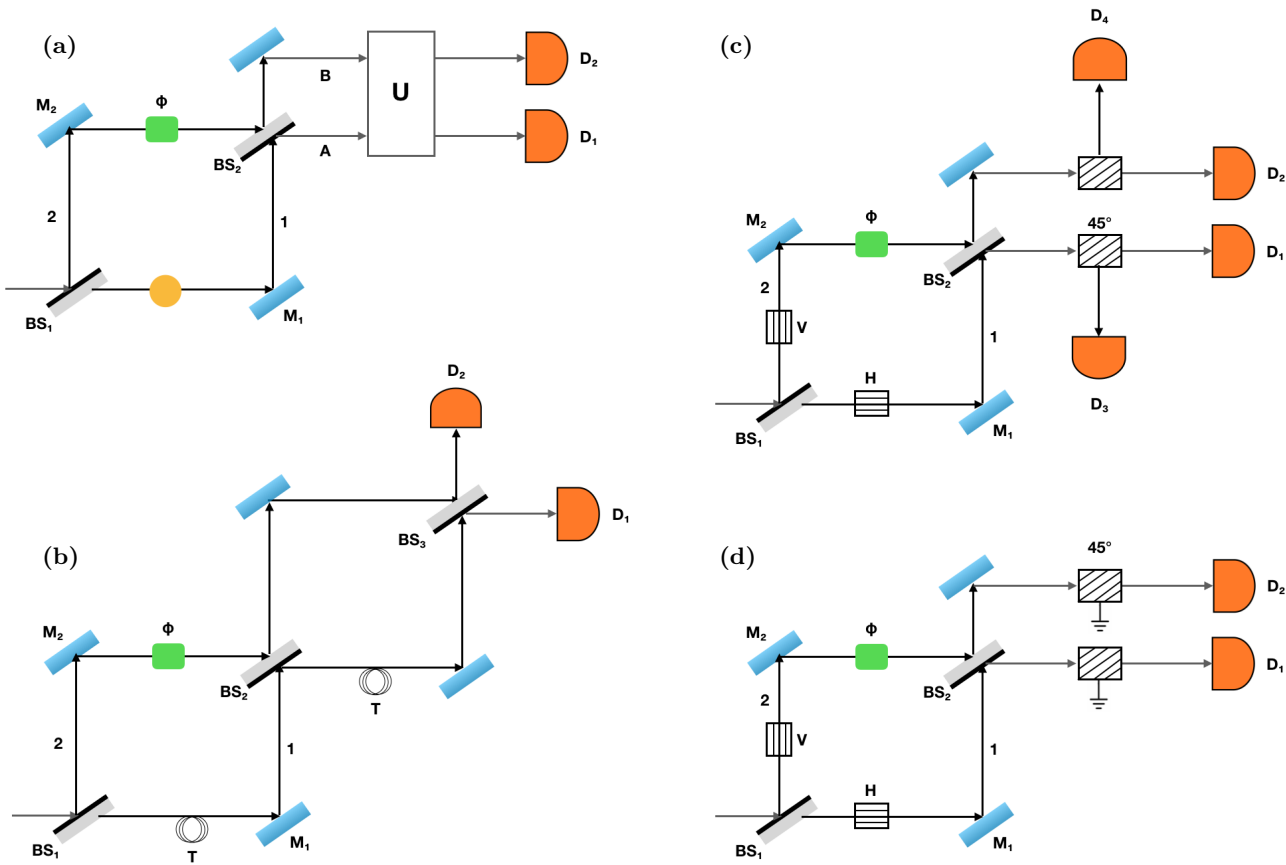
\begin{figure*}[!t] \setlength{\unitlength}{0.1cm}

\begin{picture}(110 , 110 ) 
{

\put(64, 0)  {\includegraphics[ scale=0.21]{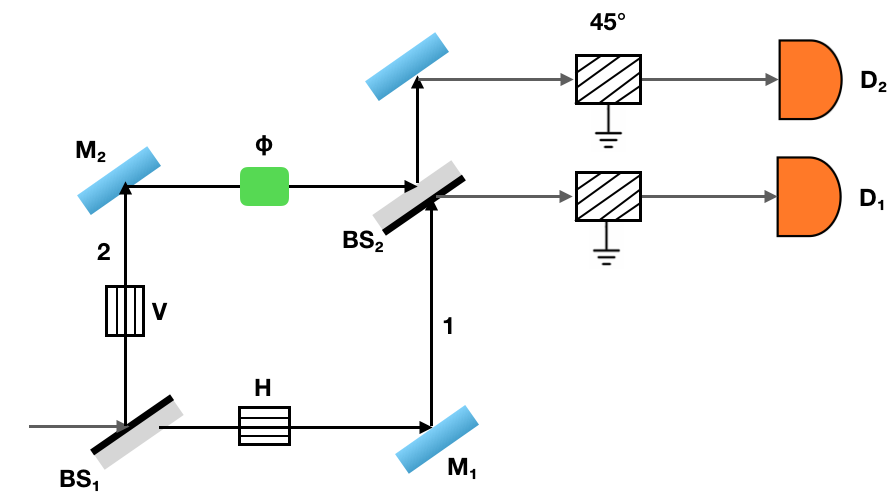}} 
\put(0, 55)  {\includegraphics[ scale=0.24]{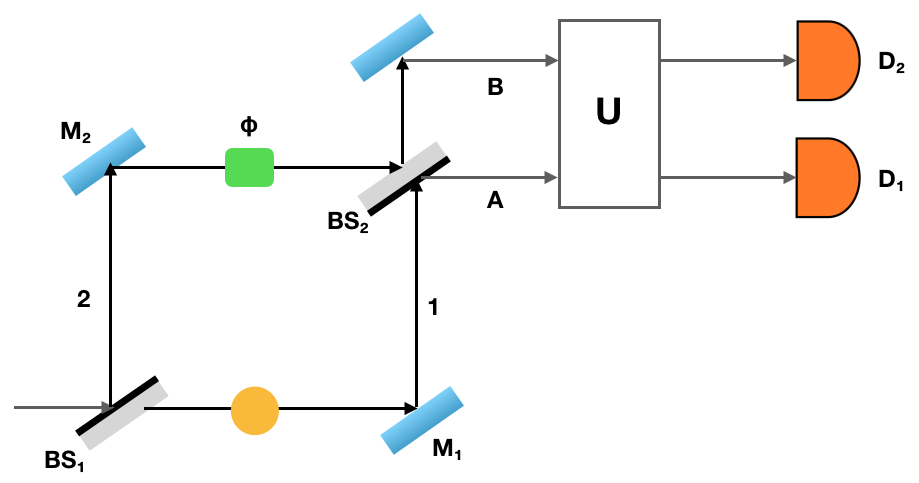}} 
\put(65, 55)  {\includegraphics[ scale=0.20]{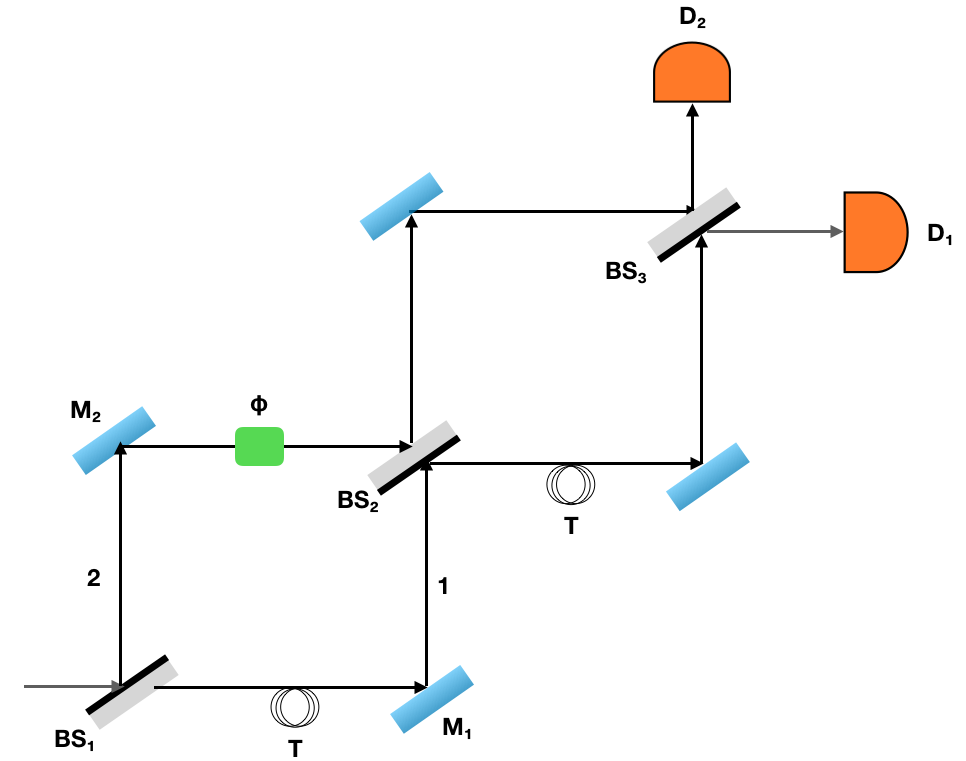}} 
\put(0, 0)  {\includegraphics[ scale=0.2]{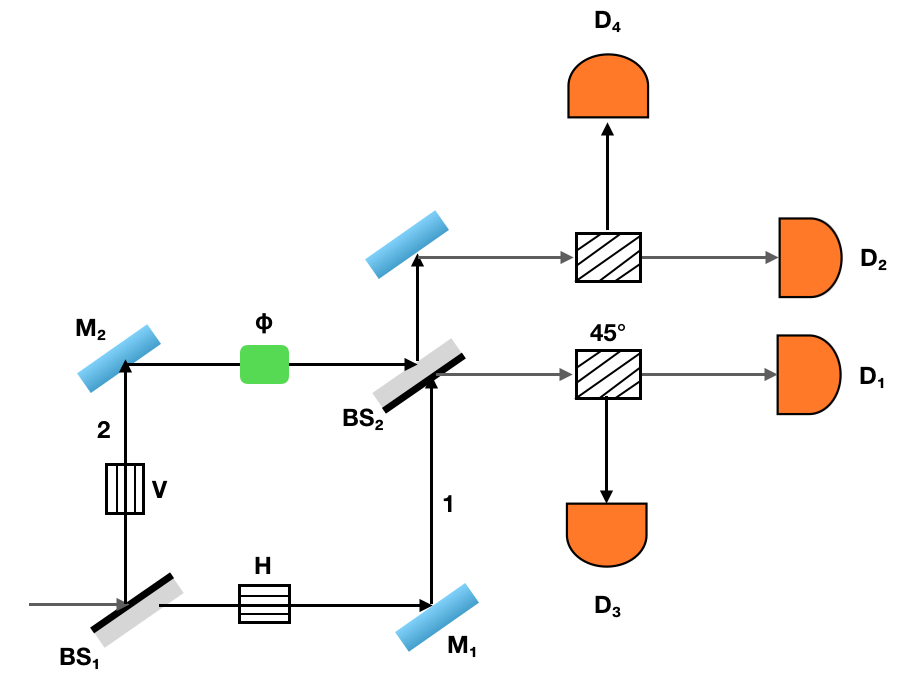}} 

\put(25,108) {\small \bf (a) }
\put(90,108) {\small \bf (b) }
\put(90,45) {\small \bf (d) }
\put(25,45) {\small \bf (c) }

}

\end{picture} 

\caption{ \label{eraser} (a) The ideal quantum eraser: A hypothetical unitary quantum circuit takes the outputs of the second beam splitter in a Mach-Zender interferometer where which-way information (WWI) is added and erases this information with a unitary transformation, thus restoring the interference pattern at $\text{D}_1$ and $\text{D}_2$ in a deterministic manner. We are not sure whether this hypothetical scheme is feasible or not.  (b,c,d) Different practical realizations of the single-photon quantum eraser. (b) Which-way information is encoded by the photon's time of arrival. 
The wavefunction of the photon arriving at the detectors at $t=T$ is in a superposition of the two paths with phase difference $\phi$ and thus produces an interference pattern. (c) Which-way information is encoded by inserting horizontal and vertical polarizers in the two paths.  Beam-splitting $45^{\circ}$ polarizers are inserted at the outputs of $\text{BS}_2$ . The wavefunctions of the photons passing through or reflected from these polarizers will be in a superposition of the two paths before they are detected by $\text{D}_1$ ($\text{D}_2$) or D3 (D4) and thus exhibit an interference pattern.  (d) Same as in (c), but using  absorbing $45^{\circ}$ polarizers after $\text{BS}_2$. Photons  will pass through the polarizers with probability 50\% and their wavefunctions will be in a superposition of the two paths, thus eliminating WWI and restoring the interference pattern at half the time.}
\end{figure*} 
 \normalsize

Let us start with the simple single-photon version of the quantum eraser. An ideal, though hypothetical, quantum erasing apparatus would utilize a unitary circuit that takes the outputs of the MZI, erases the WWI by some unitary operation and recovers the interference pattern as shown in Fig. \ref{eraser}(a). It is very likely that this ideal scheme cannot be achieved in a reversible and deterministic manner by a single unitary circuit for an arbitrary $\phi$ 
and it may be worth investigating to find a rigorous proof that this scheme does not exist. We will show in the next section an example for a unitary eraser that works for the specific values of  $\phi=0$ and $\pi$. Nevertheless, we list below three different "non-ideal" schemes,  depicted in Figs. \ref{eraser}-b, \ref{eraser}-c and \ref{eraser}-d, that erase WWI in a probabilistic manner. The main idea is to project the photon onto a state that conceals the path information or selectively measure it at times when the path information is hidden. This is accomplished by making use of EPR correlation between the two entangled degrees of freedom of the path and WWI and sorting out the registered data into different subsets that exhibit interference. 

\begin{enumerate}
\item A unitary quantum circuit is connected with the MZI that erases the WWI at certain times only in a probabilistic manner. When the WWI is encoded through the photon's time of arrival by including a delay element T in one path, the photons will arrive at $\text{BS}_2$ at $t=0$ or $t=T$ depending on the path taken by the photon. Adding another MZI with the same delay element, as in Fig. \ref{eraser}-b, will let the photons be detected at $t=0$, $t=T$ and $t=2T$ (assuming the delay associated with the  path without the delay element to be negligible). The wavefunctions of the photons detected at $t=T$ will be in a coherent superposition of  two paths with phase difference $\phi$ and thus exhibit interference. Here, we conditionally select the cases where the photon emerges in a state that conceals the path information.

\item Each of the outputs of $\text{BS}_2$  is fed into a circuit, whose two outputs exhibit the interference pattern in a complementary way. As an example, when WWI is encoded through the polarization of the photon by adding horizontal and vertical polarizers to the two paths, we can restore the interference by adding  $45^{\circ}$ polarizing beam-splitters (PBS) to the outputs of $\text{BS}_2$ (see Fig. \ref{eraser}-c). The photons emerging from each $45^{\circ}$  polarizer have either $45^{\circ}$ or $135^{\circ}$ polarization and exhibit an interference behavior. This can be easily shown by noting that the $45^{\circ}$ beam-splitting polarizers transform $|H\rangle$ and $|V\rangle$ into $\frac{1}{\sqrt{2}}(| \nearrow\rangle\pm |\nwarrow\rangle)$, where each of $|\nwarrow\rangle$ and $|\nearrow\rangle$ conceals the path information and emerges into a different detector. A simple calculation shows that the probabilities of photon detection at $\text{D}_1$ and $\text{D}_3$ (or $\text{D}_2$ and $\text{D}_4$)  are $\frac{1}{2}P_1(\phi)$ and $\frac{1}{2}P_2(\phi)$.  Note that combining the results of the two detectors in each branch yields a result equivalent to one detector that does not exhibit interference as in Fig. \ref{mzi}-b. That means that the role of the PBS is to sort the raw data that does not exhibit interference into two complementary streams of data, each exhibiting interference.

\item    Each of the outputs of $\text{BS}_2$  is fed into a dissipative circuit which randomly either absorbs the incoming photon or re-emits it with the WWI concealed. Therefore, the emerging photons will have a sinusoidal dependence on $\phi$ akin to the interference behavior without WWI. For example, considering again the MZI in Fig. \ref{mzi}-b where we encode WWI through the polarization of the photon, we can add an absorbing $45^{\circ}$ polarizer to each of the outputs of $\text{BS}_2$ (see Fig. \ref{eraser}-d). These polarizers let the photon pass through only if it is polarized at $45^{\circ}$. The wavefunctions of those photons will thus be in a coherent superposition of the two paths and will exhibit interference behavior but with a lower visibility\cite{marshman2016}. Note that in both (2) and (3), the role of the $45^{\circ}$ polarizer is to scramble the WWI encoded in the photon polarization. 

\end{enumerate}

We notice that, in case (1) above,  the second interferometer mixes the two outputs of $\text{BS}_2$ to create an interference pattern that occurs conditionally when the time delay no longer distinguishes between the two paths. On the other hand, cases (2) and (3) can be viewed as doing projective measurements in a basis that does not distinguish between the two paths.

\begin{figure*} \setlength{\unitlength}{0.1cm}

\begin{picture}(160 , 75 ) 
{
\put(13, 0)  {\includegraphics[ scale=0.45]{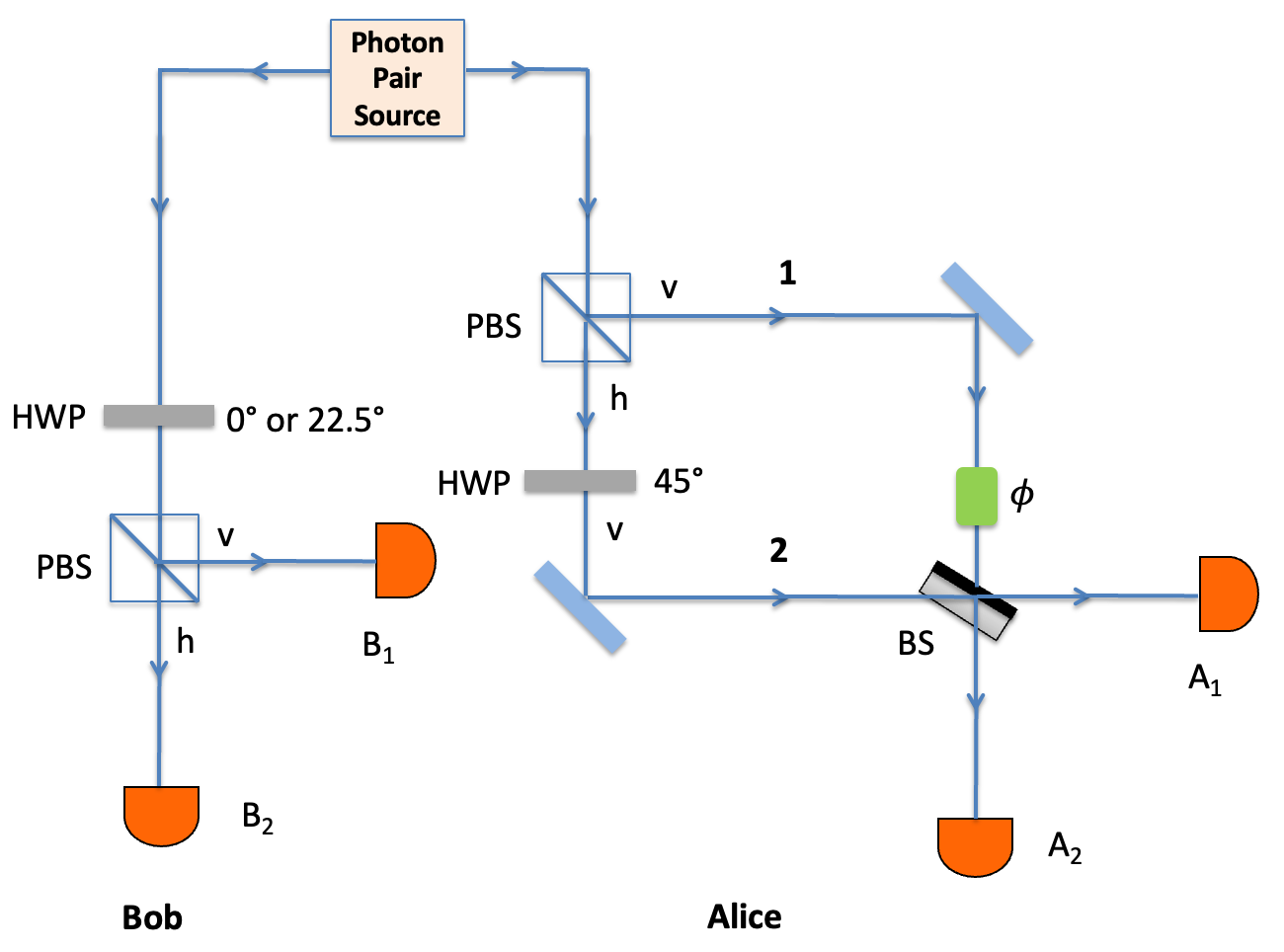}}

}
\end{picture} 

\caption{ \label{kwiat} The two-photon quantum eraser of Ref. \protect\cite{kwiat2004}. A source of entangled pairs of photons sends one photon to Alice and another photon to Bob. Depending on the measurement basis of Bob, i.e., whether he inserts or not his HWP, the probability of photon detection at Alice can exhibit an interference pattern or not.}
\end{figure*}

Now, let us turn to the two-photon quantum eraser which we adapt from Ref.\cite{kwiat2004}. A source of entangled photons generates a pair of photons in the  state $\frac{1}{\sqrt{2}}(|HV\rangle+|VH\rangle)$; the first photon is sent to Bob and the second one is sent to Alice as shown in Fig. \ref{kwiat}. Alice lets her photon enter a MZI, starting with a polarizing beam splitter that allows the vertically polarized photon to travel along path 1 while the horizontally polarized photon travels along path 2 before its polarization direction is flipped to the vertical direction by a 45° HWP. Alice further introduces a phase shift $\phi$ along path 1, making the full wavefunction of the two photons at this point $\Psi=\frac{1}{\sqrt{2}}(|V2\rangle+e^{i\phi}|H1\rangle)$. If Bob measures his photon in the HV basis, his measurement outcome will serve as WWI for Alices's photon, thus removing any dependence in the probability of detection at $\text{A}_1$ and $\text{A}_2$ on $\phi$. On the other hand, if Bob measures his photon in the rectilinear basis $\psi_\pm=\frac{1}{\sqrt{2}}(|H\rangle\pm|V\rangle)$ by introducing a 22.5° HWP before his PBS, the path information of Alice's photon will be eliminated. To see this, let us compute the state of Alice's photon after the detection of  Bob's photon by applying the projection operator $|\psi_\pm\rangle \langle \psi_\pm|$ to $\Psi$, i.e., $|\psi_\pm\rangle \langle \psi_\pm| \Psi\rangle=\frac{1}{2}(|H\rangle\pm|V\rangle)\otimes (e^{i\phi}|1\rangle\pm|2\rangle)$. As shown before, this result makes the probability of photon's detection by Alice at $\text{A}_1$ to be either $P_1(\phi)$ or $P_2(\phi)$ depending on the measurement outcome of Bob. We emphasize here that the operation of this setup relies on the EPR correlation between the two particles which is a consequence of   them being in an entangled quantum state.

 \normalsize

\section{A QKD algorithm based on the quantum eraser}
We are now ready to present a scheme for securely sharing a random key between two agents using the quantum eraser. The aim is to let Alice generate a key consisting of logical bits 1 and 0 and securely share it with Bob, or the other way around, in order to use it for encrypting messages. The key should be communicated such that any eavesdropping attempt by Evan, an evil agent trying to intercept the message, can be detected by either Alice or Bob.

In the BB84 algorithm, which uses single photons, the sender chooses randomly either one of two different basis states of the photon polarization, i.e., rectilinear or diagonal basis. The logical bit is encoded through two  perpendicular polarization directions corresponding to bits 1 and 0. The receiver, on the other hand, chooses randomly to measure the state of the photon in any of the two bases. The sender and receiver  will communicate in public their basis choices and register the corresponding bits only when their choices agree. In doing so, they generate a shared random key that they can use later for encrypting a message sent on the public channel. By comparing  a subset of the generated data, they can detect any attempt of eavesdropping on the communication since any intervention from a third party in the middle will eventually lead to some errors in the generated data. The security of this protocol has been analyzed by many authors (see e.g.,\cite{shor2000,mayers2001,lo2014}).

On the other hand, in the Ekert91 algorithm, a source generating a stream of entangled pairs of photons sends one photon from each pair to Alice and another one to Bob. Each of them has the freedom to choose the measurement basis and at the end of the transmission they announce their basis  in public. They also perform an entanglement check for a random subset of the photons to detect eavesdropping, i.e., to make sure that the pairs are entangled by using Bell's inequality measurement.  In principle, the two schemes are similar, except for the security check step performed by the two parties.

Here, we aim at using the quantum eraser as a platform for generating the key.  The main idea is to let one party control a MZI and another party control the quantum eraser and make use of the various degrees of freedom available at the two parties related to inserting the WWI or erasing it to generate a shared key.
Consider first the schemes of Fig. \ref{eraser} and let Alice use an MZI to encode the logical bits by the state of the photon. She has two degrees of freedom, the phase difference between the two paths and whether to include WWI or not. Let  $\phi=\pi$ corresponds to bit 1 and  $\phi=0$ corresponds to bit 0. The two outputs of $\text{BS}_2$  are sent to Bob. Bob, on his part, has the freedom to insert a quantum eraser or not. 
In order to share a random secret key, Alice sends a stream of photons to Bob, with randomly setting  $\phi=\pi$ or $\phi=0$ while at the same time randomly choosing to insert WWI or not. Bob, on the other side, randomly chooses to insert the eraser setup or not and records through which detector the photon was detected. At the end of the transmission, Alice and Bob announce over the public channel their choices of WWI and the eraser setup for each transmission event. Bob also announces whether his eraser setup successfully erased the WWI, i.e., the photon was detected at the right time in Fig. \ref{eraser}-b or was actually detected in either $\text{D}_1$ or $\text{D}_2$ in Fig. \ref{eraser}-d. Unlike the schemes in Figs. \ref{eraser}-b  and \ref{eraser}-d, the scheme in Fig. \ref{eraser}-c is lossless, i.e., the photons always emerge in either of the two branches creating an interference of $P_1(\phi)$ or  $P_2(\phi)$. In this scheme, Bob should be careful with the bookkeeping of the detected photons and use the respective rule with each branch; a photon's detection at $\text{D}_1$ and $\text{D}_4$ corresponds to logical bit 0, while a photon's detection at $\text{D}_2$ and $\text{D}_3$ corresponds to logical bit 1.

A logical bit will be securely shared between Alice and Bob  when Alice inserts WWI and Bob utilizes the quantum eraser and successfully registers the photon or when Alice does not insert WWI and Bob does not insert the quantum  eraser. In these two cases, Bob knows with certainty the value of $\phi$ used by Alice through the measurement outcomes of his detectors. The bits corresponding to these cases are kept by Alice and Bob as parts of the key. 
According to the no-cloning theorem\cite{wootters1982}, Evan can not measure the phase relationship between the two channels and the state of WWI (i.e., whether it was inserted or not and what it was) simultaneously without disturbing the state of the photon in the channels, thus causing a discrepancy between the values of the bits registered by Alice and Bob. Therefore, Alice and Bob can detect the existence of an eavesdropper by comparing over the public channel a subset of their data that will be discarded later.

It is evident that this algorithm is conceptually identical to the BB84 algorithm. Therefore, the security of our scheme can be arbitrarily improved using privacy amplification methods\cite{bennett1992,bennett1995}. On the other hand, in addition to being quite impractical and more complicated than BB84,  it suffers from the main drawback of the BB84\cite{lo2014}, namely that it requires  very low-noise detectors and efficient single-photon sources. The reason for the necessity of single-photon sources is that inefficient single-photon generation will send multiple photons in the same quantum state, allowing Evan to secretly intercept one of them.

Now, let us consider the two-photon quantum eraser of Fig. \ref{kwiat}. Basically, Bob will decide randomly to insert his 22.5° HWP or not, while Alice will decide randomly to use  $\phi=0$ or $\phi=\pi$. This choice of $\phi$ makes each of $P_1(\phi)$ and $P_2(\phi)$ either 0 or 1. Alice can thus know  the probability rule imposed on her by the measurement outcome at Bob's side, should he choose to erase WWI by inserting the 22.5° HWP, once the photon is detected at either $\text{A}_1$ or $\text{A}_2$. At the end of the transmission of many pairs of photons, Bob will communicate his choices to Alice, and at the instances he decided to insert the eraser, she will know whether his photon was detected at $\text{B}_1$ or $\text{B}_2$ by computing the probability rule governing her measurement outcomes by comparing the value of $\phi$ and the detector at which her photon was detected (either $\text{A}_1$ or $\text{A}_2$). A shared qubit will be registered as 1 or 0 when Bob inserts the erasing HWP depending on whether the photon is measured in $\text{B}_1$ or $\text{B}_2$


 \begin{figure*}[!t] \setlength{\unitlength}{0.1cm}

\begin{picture}(160 , 85 ) 
{

\put(0, 0)  {\includegraphics[ scale=0.4]{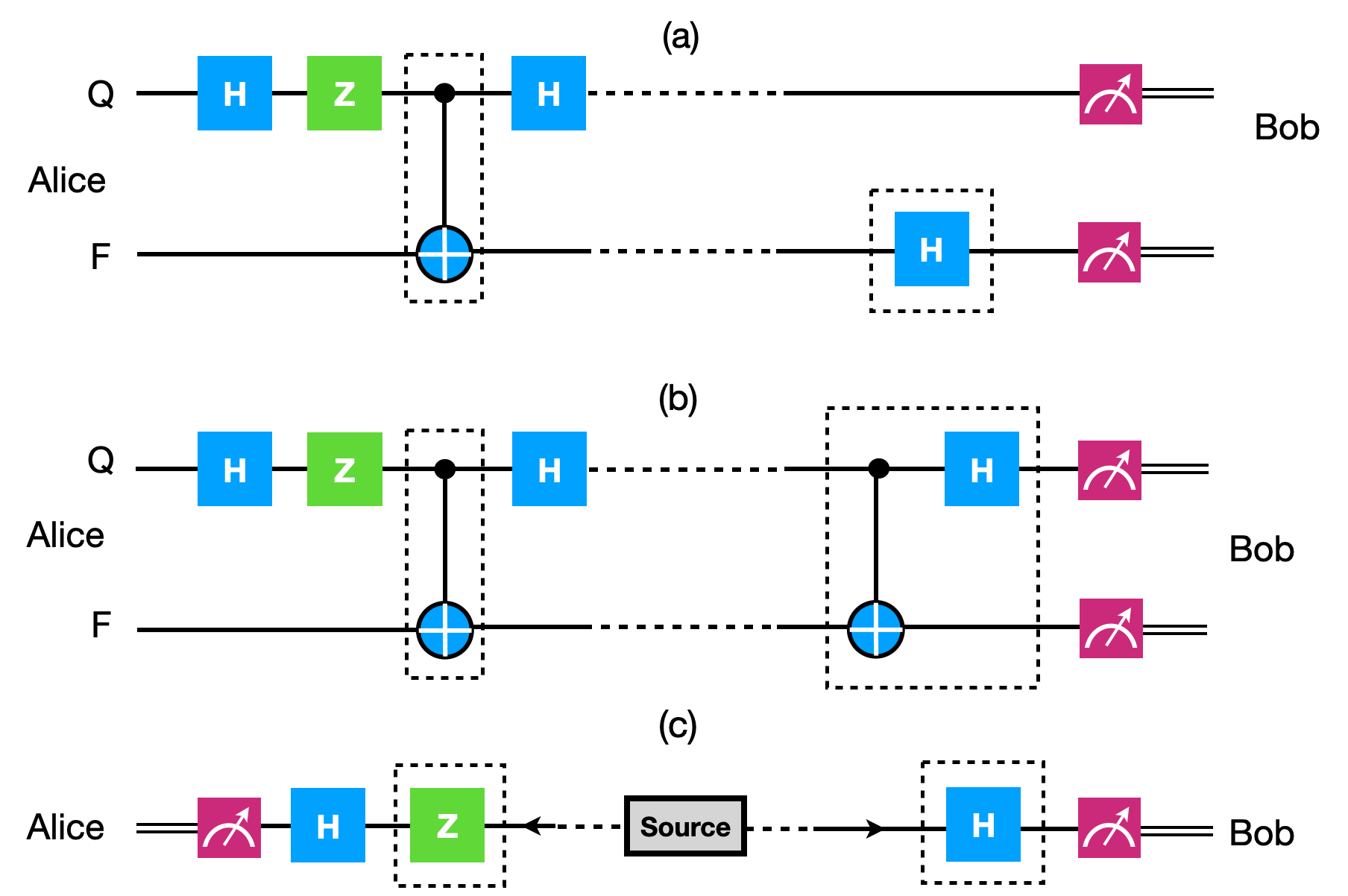}}

}
\end{picture} 

\caption{ \label{circuit} Quantum circuit implementations for the QKD schemes based on the quantum eraser. (a,b) Alice sends two qubits Q and F to Bob and chooses randomly to insert the CNOT and Z  gates or not. The Z gate determines whether bit 1 or 0  is sent, while the CNOT gate determines whether which-way information (WWI) is sent or not. Bob measures the state of the two qubits after choosing randomly to insert the eraser circuit or not. (c) A source of entangled qubits sends one qubit to Alice and another one to Bob.  The dotted boxes in the three figures indicate the blocks that are randomly inserted by each side to include WWI or the quantum eraser. }
\end{figure*} 

 \normalsize

\section{Quantum circuit implementation}

In this section, we introduce a quantum circuit implementation of the proposed schemes using basic quantum logic gates. First, consider the schemes based on the single photon quantum eraser. Alice sends two qubits, Q and F, to Bob as shown in Fig. \ref{circuit}-a and \ref{circuit}-b that are both initialized to $|0\rangle$. Q represents the  photon in the quantum eraser while the flag qubit F represents the which-way information. The states $|0\rangle$ and $|1\rangle$ of Q represent the two different paths of the photon inside the MZI. Note that in the MZI of section 2, WWI is encoded through additional degrees of freedom of the same photon, while here we encode WWI through another qubit. The two Hadamard gates (H) in Fig. \ref{circuit}-a and \ref{circuit}-b at the side of Alice represent the two beam splitters while the Z gate represents the phase element $\phi=\pi$. When Alice chooses to insert WWI, she inserts a controlled-NOT (CNOT) gate between the two Hadmard gates that entangles the two qubits. On the other side, before Bob performs his measurements in the Z-basis on the two qubits received from Alice he can choose to erase the WWI or not. We show two different implementations for how Bob can erase the WWI in Fig. \ref{circuit}-a and \ref{circuit}-b. When he decides to erase the WWI, Bob can either insert an H gate to F before he performs the measurement as in Fig.  \ref{circuit}-a, thus effectively measuring it in the X-basis, or  alternatively,  he can insert a CNOT gate to F followed by a Hadmard gate to Q before doing the measurements in the Z-basis as in Fig.  \ref{circuit}-b. In both cases, the information about the WWI encoded in the Z-basis is scrambled.

Let us analyze the operation of the two eraser circuits in Fig. \ref{circuit}-a and \ref{circuit}-b. When Alice and Bob compare their choices at the end of the transmission, the simplest scenario will correspond to the case when Alice chooses not to insert WWI and Bob chooses not to insert the eraser. In this case,  Bob can detect directly whether Z-gate was inserted at Alice's side or not  by measuring Q alone since $\text{H}^2|0\rangle=|0\rangle$ and $\text{HZH}|0\rangle=|1\rangle$. The F qubit is redundant in this case and therefore Alice can always set it to a random value whenever she does not insert WWI. The other scenario corresponding to the valid generation of a secretly shared bit occurs when Alice chooses to insert WWI and Bob chooses to insert the quantum eraser.   Starting with $|\text{QF}\rangle=|00\rangle$, it is straightforward to see that with the Z-gate inserted, the state of the two qubits at Bob's side directly before the measurement in  Fig. \ref{circuit}-a is 
$\frac{1}{\sqrt{2}}(|01\rangle+|10\rangle)$
and without the Z-gate is $\frac{1}{\sqrt{2}}(|00\rangle+|11\rangle)$,
while in Fig. \ref{circuit}-b it
 is $\frac{1}{\sqrt{2}}(|00\rangle-|11\rangle)$ with the Z-gate inserted and $\frac{1}{\sqrt{2}}(|10\rangle+|01\rangle)$ without the Z-gate. Bob can thus detect whether the Z-gate is inserted or not while using either of the two circuits by detecting whether the measurements of Q and F yield similar or different results. If Z-gate is inserted, bit 1 is recorded and if Z-gate is not inserted, bit 0 is recorded. As in BB84, Alice and Bob will communicate their choices over the classical channel to determine the cases where their choices comply with each other.

Unlike the BB84, which can be implemented with a single qubit, we need here two qubits to implement this protocol, one of them is redundant half the time. The redundancy of the flag qubit can be reduced by several methods. For example, by using the flag bit to send another bit encoded in the Z-basis that Bob will record as is during the No-WWI-No-Eraser mode, therefore sharing two bits instead of one bit during that mode. If Evan decides to intercept this bit, he will eventually introduce errors during the WWI-Eraser mode since he does not know a priori when each mode occurs. Another example is to send two qubits $\text{Q}_1$ and $\text{Q}_2$ served by the same F qubit, thus doubling the transmission rate. In this case, Alice and Bob use WWI-Eraser mode with one of the two qubits (selected randomly by each of them every time) and the other mode with the other qubit. As before, two secret bits will be shared when the choices of Alice and Bob are consistent. It is interesting to note that the two circuits in Fig. \ref{circuit}-a and \ref{circuit}-b used to erase WWI are completely different; in one case Bob treats the two qubits independently while in the other case he uses an entangling two-qubit operation that acts on Q and F before doing the measurement.  

One more difference between the schemes in Figs. \ref{circuit}-a and \ref{circuit}-b on one hand and conventional QKD algorithms such as BB84 on the other hand is the distinguishability between the different basis states that are used for representing the logical bits. The closer these  basis states are to each other, the harder it is for Eve to distinguish between them. In BB84, for example, the overlap between the two bases is $\frac{1}{2}$. In the schemes in Fig. \ref{circuit}-a and \ref{circuit}-b, the states sent over the channel for logical bit 0 is $|00\rangle$ when no WWI is sent and $\frac{1}{2}(|00\rangle+|01\rangle+|10\rangle-|11\rangle)$ when WWI is sent. The overlap between the two states is $\frac{1}{4}$. Therefore, they are more distinguishable than the BB84 algorithm and that makes the scheme less efficient.
In the two schemes, quantum degrees of freedom are used to generate classical information. In the BB84, the two parties freely choose the measurement basis of the photons while, here, they freely choose to insert or erase the which-way information.   In BB84, single-qubit transmission is used where both sides can choose whether or not to insert a Hadamard gate before the transmission or the measurement process. Here, the path information is encoded through another qubit that is entangled with the first one. This reminds us of the deterministic quantum protocols which require at least four degrees of freedom for securely sending one deterministic bit using a one-way quantum channel\cite{beige2001,beige2002}.

Let us now consider a quantum circuit corresponding to the scheme based on the two-photon eraser. As depicted in Fig. \ref{circuit}-c, a source of entangled qubits generates pairs of qubits in the state $\frac{1}{\sqrt{2}}(|01\rangle+|10\rangle)$ and sends one qubit to Alice and another one to Bob. Bob has the freedom to insert a Hadamard gate before the measurement, which corresponds to inserting the 22.5° HWP element that erases WWI. On the other hand, Alice has the freedom to insert her Z gate, which corresponds to setting $\phi=\pi$ in Fig. \ref{kwiat}, before she lets her qubit go through a Hadamard gate. A new bit of the key will be registered only when the eraser element is inserted by Bob, which will preserve the correlation between the measurement outcomes of both sides. The measurement outcomes of Alice and Bob will be identical if Z-gate is not inserted, and will be the opposite of each other when Z-gate is inserted. Alice and Bob will agree a priori on whose version will be used in the key in the later case. At the end of the transmission, Alice and Bob will communicate their choices in public and thus a random and secure key will be shared between them.

In a sense, this protocol is quite similar to Ekert91\cite{ekert91} protocol where a source of maximally entangled pairs of photons sends one photon to Alice and another one to Bob and each one chooses the measurement basis randomly. Here, the protocol was motivated by and conceived of as a direct mapping of the quantum eraser scheme of\cite{kwiat2004}. An eavesdropper trying to hijack the communication would intercept the original pair of qubits and send instead a pair of disentangled qubits, each in a certain state.  As in other schemes that use entangled pairs of photons such as Ekert91, a Bell's test should be conducted by Alice and Bob on a subset of the pairs in order to verify that the states of the two qubit are maximally entangled Bell states and thus detect the presence of an eavesdropper.

\section{Discussion and Conclusion}

Let us analyze the most famous eavesdropping attack strategy, namely the intercept and resend attack, and compute the quantum bit error rate (QBER) for the quantum circuits in Fig. \ref{circuit}-a and \ref{circuit}-b. In this attack, Evan poses as Bob, intercepts the two qubits, makes measurements with or without an eraser setup, and sends a new version of the two qubits forward to Bob. Let us first consider the simple  No-WWI-No-Eraser mode. In this case, the two qubits are not entangled, and  the flag qubit has a random value. Suppose that Evan measures Q and resends it to Bob, thus always obtaining a valid bit in this mode which occurs 50\% of the time. Since Evan does not know beforehand when this mode occurs, he will inadvertently corrupt the two-qubit state during the WWI-Eraser mode. More specifically, since the two-qubit state in the channel during this mode is an equal-weighted superposition of all the basis states in the Hilbert space, the F qubit will be projected due to Evan's intervention to a superposition of $|0\rangle$ and $1\rangle$. When Bob inserts his eraser circuit, the full state will eventually be an equal mixture of all the four basis $\{|00\rangle,\ |01\rangle,\ |10\rangle, \  |11\rangle \}$ regardless of whether 0 or 1 was encoded by Alice. Therefore, unsurprisingly like the BB84 algorithm, Evan will corrupt on average 25\% of the bits of the shared key, i.e., QBER=25\%. Similarly, if Evan chooses to intercept both Q and F with an eraser circuit, posing as Bob, and then generates an entangled state and sends it to Bob, posing as Alice, he will inadvertently corrupt  25\%  of all the shared bits.

While the algorithms presented in this paper may not be efficient from the practical point of view, they are useful from the pedagogical point of view. The QKD algorithms presented above show that quantum communication algorithms can be motivated using notions in the foundations of quantum mechanics such as the quantum eraser. Although perhaps overcomplicated, the proposed circuits serve as examples of emulating a quantum interference experiment (the quantum eraser) using elements of quantum circuits.

In conclusion, it was shown in this paper how closely the fundamental notions of quantum mechanics  and the emerging field of quantum algorithms and quantum information are connected. This connection exists because foundational issues in quantum mechanics are best explored with experiments performed on single quantum particles, while qubits, the basic unit of quantum information, are carried by single quantum particles as well, i.e., single photons.   As such, the two fields are often overlapping. For example, entanglement which is a key concept in the foundation of quantum mechanics is one of the most important resources in quantum information theory. Given that much of the fascination of students with quantum mechanics comes from the paradoxes and controversial issues in quantum foundations, bringing them in touch with this field early on in their education may lead to an accelerated progress in the field of quantum foundations and more likely in the emerging field of quantum technologies.

\subsection*{Acknowledgment} 
The author thanks Prof. Mark Hillery for the many discussions, his comments on the paper and for the hospitality of Hunter College of CUNY where this work was conceived. The author thanks also Prof. Daniel Greenberger for the discussion.

\bibliographystyle{ws-ijqi}

\end{document}